\documentclass{article}

\usepackage{neurips_2019}

\usepackage[utf8]{inputenc} 
\usepackage[T1]{fontenc}    
\usepackage{hyperref}       
\usepackage{url}            
\usepackage{booktabs}       
\usepackage{amsfonts}       
\usepackage{nicefrac}       
\usepackage{microtype}      
\usepackage{graphicx}
\usepackage{url}
\usepackage{float}
\graphicspath{{./Images/}}
\setcitestyle{numbers}

\title{A Virtual Reality Game as a Tool to Assess Physiological Correlations of Stress}

\author{
  Daniel Lee\\
}

\begin{document}

\maketitle

\begin{abstract}

The objective of this study is to develop and use a virtual reality game as a tool to assess the effects of realistic stress on the behavioral and physiological responses of participants. The game is based on a popular Steam game called Keep Talking Nobody Explodes, where the player collaborates with another person to defuse a bomb. Varying levels of difficulties in solving a puzzle and time pressures will result in different stress levels that can be measured in terms of errors, response time lengths, and other physiological measurements. The game was developed using 3D programming tools including Blender and virtual reality development kit (VRTK). To measure response times accurately, we added LSL (Lab Stream Layer) Markers to collect and synchronize physiological signals, behavioral data, and the timing of game events. We recorded Electrocardiogram (ECG) data during gameplay to assess heart rate and heart-rate variability (HRV) that have been shown as reliable indicators of stress. Our empirical results showed that heart rate increased significantly while HRV reduced significantly when the participants under high stress, which are consistent with the prior mainstream stress research. We further experimented with other tools to enhance communication between two players under adverse conditions and found that an automatic speech recognition software effectively enhanced the communication between the players by displaying keywords into the player's headset that lead to the facilitation of finding the solution of the puzzles or modules. This VR game framework is publicly available in Github and allows researchers to measure and synchronize other physiological signals such as electroencephalogram, electromyogram, and pupillometry. 

\end{abstract}

\section{Introduction}
Various real-life stress can induce homeostatic changes in human behavior, brain, and body. For instance, associations have been reported between achievement on tests of memory and attention and the experience of every day minor stressful events\cite{Neupert2006, Stawski2011}. 
To modify the amount of stress that a person has, researchers have been relying on manipulating the various factors of their subjects’ lives, such as their sleep quality and workload. However, it is unrealistic to control every aspect of the subjects’ lives, and they may always be more or less stressed than the researchers think they are, varying the data from the experiment. 
Furthermore, stress research to date centers on simplified stimulus-response paradigms conducted in highly controlled environments that do not resemble authentic settings, where real-life stress typically takes place. We propose developing a controlled virtual-reality (VR) environment that can allow for greater control over the subjects’ stress and performance when completing tasks\cite{bailenson_2019}. These types of games are known as games with a purpose (GWAP). They are tools for helping scientists perform experiments and collect data \cite{lance}. 

\subsection{Games With A Purpose}
There are many examples of GWAP, and most of them were designed to collect data from a large player base to build on research or improve programs\cite{lance}. An example of such a game is the Airport Scanner game, which makes players find airport contraband using an X-ray scanner. This data set was used by researchers to investigate how well baggage screening can be performed.
Games have also been created for studying performance in cognitive tasks. An example of this is Luminosity, where players play simple games that train various aspects of cognition. Scientists hope to use this database to gain better insights into how people think and what factors affect cognitive ability. In addition, games like Foldit \cite{cooper} or Eyewire \cite{Eyewire} use players to advance scientific studies, such as folding complex proteins or mapping out the synapses of the brain, respectively. Using these games while adding neural and physiological measurements is expected to introduce a new dimension to user interaction and monitoring experiments \cite{10.1007/978-3-319-39955-3_12}.

This study modified a popular VR game, Keep Talking and Nobody Explodes \cite{keep_talking}, to test participants in a controlled environment. Keep Talking Nobody Explodes is a VR game where one player disarms a bomb while communicating with another player, a bomb-defusing expert, who has the bomb defusing instructions. The modules or puzzles for disarming the bomb include certain wires that must be cut, buttons that must be pressed at certain times, and keypads that must be pressed in a certain order. The player will need a VR headset (e.g. Oculus or HTC Vive) and controllers to play the game.  Only the player can see the puzzles and bomb in VR, while the bomb-defusing expert can only check a manual to find instructions to solve the puzzles. The player and the bomb-defusing expert must communicate effectively to defuse the bomb.

\subsection{Research Goals of our GWAP}
The ultimate goal of this research is to find methods to reduce stress and improve performance at a certain task. The present study aims to find physiological markers under stressful and harsh conditions relating to potential real-life situations. Finding ways to measure stress and how it affects performance is difficult in the way that it is difficult to control a subject's stress level. This study develops a VR game framework that can be used to measure and synchronize physiological signals, behavioral data, and the timing of game events. The framework needs to be flexible enough to expand to multiple data modalities and it needs to be able to include tools to assist in solving the tasks. In essence, this game was meant to be a different version of an already existing game meant for entertainment. It was a recreation of an existing game and re-purposed as a game with a scientific purpose so that researchers could use this game as a tool for studying stress in-depth in a controlled environment. Furthermore, this game needs to be enjoyable enough so that its players are willing to return to play the game with different parameters multiple times. The scientists then can acquire sufficient data over different days/times to allow the researchers to study the variability in the stress-performance associations across different individuals and times within an individual.

\section{Methodology}

To create a controlled environment where researchers can measure the performance of a task and manipulate how well the subject performs, we designed a VR game based on the game Keep Talking and Nobody Explodes \cite{keep_talking}, which measures the subjects’ reaction times, communication times, and precision and recall of tasks. There are multiple levels of difficulty expressed in more puzzles (modules) and less time to solve them.  There are many ways to measure stress and among those one of the most widely studied methods is to measure heart rate and HRV derived from ECG recordings \cite{Stress_HRV}. We developed a simple framework where we can record the ECG signals while playing the game under varying stress conditions and mark the ECG signals with start and end times of solving a game module or events that are triggered within the game. The event markers are created by the Lab Streaming Layer (LSL), a recording and annotation software \cite{LSL} that synchronizes the recorded signals with events during the game. 

Our method is also extendable with the inclusion of other physiological data such as Electroencephalogram (EEG), electromyogram (EMG), and pupilometry. Similar to ECG, the EEG/EMG data can be integrated with this game through the LSL software.

\subsection{Overview of the VR Game Design of Keep Talking and Nobody Explodes }

The game we created originates from a popular social game called Keep Talking and Nobody Explodes, where one player disarms a bomb based on another player’s instructions. The player giving the instructions has a bomb-defusing manual that has instructions to solving each puzzle module, and when all the puzzle modules are solved, the bomb is disarmed. The player who has the instructions, or the bomb-defusing expert, cannot see the bomb or the puzzles, and the player who can see the bomb does not have the instructions. Therefore, the two players must collaborate and communicate effectively to disarm the bomb.

\begin{figure}[h]
  \centering
  \includegraphics[height=6cm]{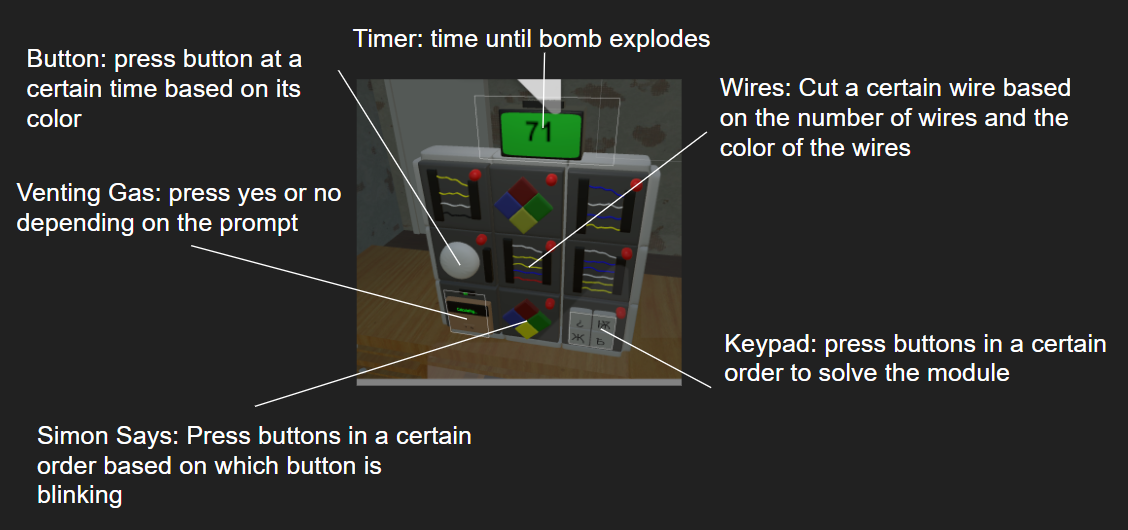}.
  \caption{An image of the bomb in the VR game. The top left, right and middle portions of the bomb are the wire modules. The top middle and bottom middle modules are the Simon Says modules. The solution manual is depicted in Figure~\ref{bomb_manual} (A). The module in the bottom right is the keypad module, the module in the middle left is the button module, and the module in the bottom left is the venting gas module. The solution to the keypad module is depicted in Figure~\ref{bomb_manual} (B).}
  \label{Bomb_Image}
\end{figure}

In our version of the game, we implemented the following five types of modules: Wires, Keypad, Simon Says, Button and Venting Gas. Figure~\ref{Bomb_Image} shows the image of the bomb with the five types of modules. The timer at the top of the bomb indicates the time before the bomb explodes, and how many mistakes the player has made (marked as an x per mistake, which is not shown in the figure). The timer was implemented using the Unity time.time function. For the Wire module (shown in the top left, top right, and two in the middle right), the player is required to cut a certain wire based on the color of the wires and how many there are. For the Keypad module (shown at the bottom right), the player must press the keypad in a certain order depending on the symbols displayed on the keypad. For the Simon Says module (shown in the top and bottom middle), the player must press the buttons in a certain order based on which button is blinking. For the Button module (shown in the leftmost middle row), the player must press the button at a certain time to solve it. The Venting Gas module, in the bottom left, is a “needy module” that captures the player's attention and causes more stress.

\begin{figure}[h]
  \centering
  \includegraphics[height=5cm]{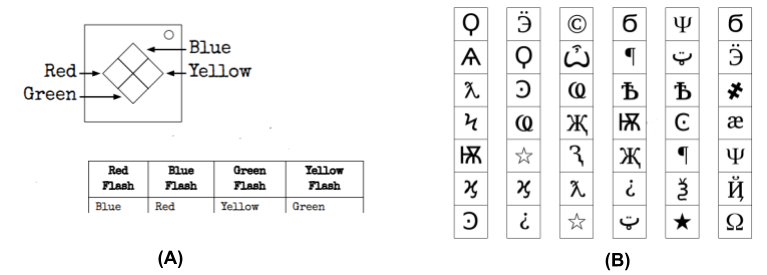}.
  \caption{Two sample pages of the bomb manual used in this study. The purpose is to allow the bomb-defusing expert to explain to the player how to solve the puzzles in the game. (A) the bomb-defusing instructions for the Simon Says module; (B) the bomb-defusing instructions for the keypad.}
  \label{bomb_manual}
\end{figure}

These modules are solved based on our version of the bomb-defusing manual, part of which is shown in Figure~\ref{bomb_manual}. When playing this game, one player with the VR headset can see the bomb, as shown in Figure~\ref{Bomb_Image}, while the other player reads this manual to instruct the player who is disarming the bomb. 

In our initial implementation, there were three difficulty levels: easy, with four modules to solve in five minutes, medium, with six modules to solve in two minutes, and hard, with nine modules to solve in one and a half minutes, as shown in Table~\ref{difficulty_table}. 
Because this study focuses on the physiological correlates (e.g. HR and HRV) of stress levels (easy vs hard), we simplified the difficulty levels of our GWAP so that the players were instructed to solve a random puzzle in two minutes  (easy) or twenty seconds (hard) to simulate non-stressful and stressful conditions, respectively. We describe this experiment and its results in more detail in the following sections.

\begin{table}[h]
\centering
\caption {Difficulty Table for our initial version of Keep Talking and Nobody Explodes}
\begin{tabular}{llll}
\hline
\textbf{Difficulty}    & \textbf{Easy}  & \textbf{Medium}    & \textbf{Hard}        \\ \hline
\textbf{Time}          & 5 min & 2 min & 1.5 min \\ \hline
\textbf{Number of Modules} & 4         & 6         & 9\\ \hline
\end{tabular}
\label{difficulty_table}
\end{table}

\subsection{Software Tools Used to Develop the VR Game}
\label{gen_inst}

We used Unity \cite{unity}, a popular game engine tool because it uses a well-known programming language and has a simple UI. We also used Blender \cite{foundation}, an open-source rendering tool, because of its popularity and vast amounts of resources online.

We made 3D models such as the bomb itself in blender and imported them into Unity. Afterward, the textures were applied to the game objects in Unity. After applying these textures found from the Unity store and online, we started to create the codes in the game. We first implemented the five types of modules (Wire, Button, Simon Says, Keypad, and Venting Gas, which are described in the previous section) in C\#.

To convert our game into VR, we used VR Toolkit \cite{VRTK}, a VR library compatible with Unity. Some of the functions (such as grabbing an object and moving it) were already included in the VR library, but we had to modify others so that our game would work with those VR functions that allowed the player to use the game objects.

\subsection{Adding Event Markers into the VR GWAP for Time Measurements}
Lab Streaming Layer (LSL) \cite{LSL} is a scientific tool used to synchronize streaming data for live analysis or recording. We used LSL Markers to indicate when certain events triggered, like when the player pressed a specific button or cut a specific wire. LSL is a cost-efficient data acquisition and synchronization framework for distributed systems with a focus on reliability, near-real-time access, and time-synchronization. We used the Unity LSL library to mark and broadcast certain events and output timestamps with those events. 
In our GWAP game, the event markers included when the player started each session (resting, hard, easy sessions), when the player pressed a button or cut a wire, when the player solved a module, and when each session ended.

\begin{figure}[H]
  \centering
  \includegraphics[height=6cm]{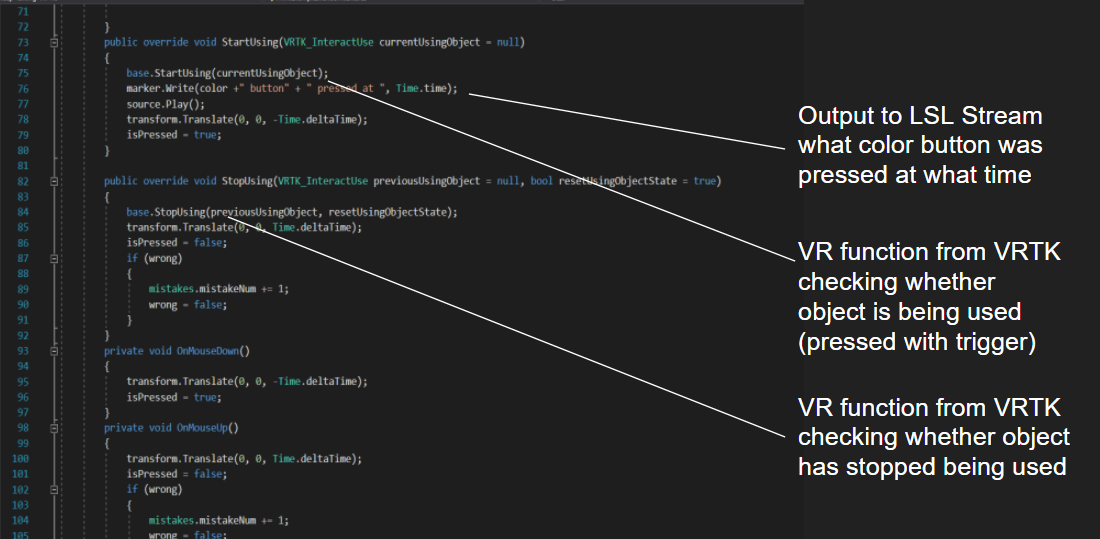}.
  \caption{The code used for outputting an event marker and using VR functions like grabbing and moving an object.}
  \label{LSL_Code}
\end{figure}

Figure~\ref{LSL_Code} shows a sample code used to output an event marker when the button module is pressed. Lines 73 to 79 define the function \texttt{StartUsing}, which is called when the button is pressed in VR. Line 76 is where the event marker itself is sent out. Lines 82 to 91 define the function \texttt{StopUsing}, which is called when the button is released and checks whether the player has solved the module or not. The rest of the codes can be viewed through the GitHub link \cite{lee_xu}.

\subsection{Recording and Analyzing Electrocardiogram Data}

This section presents how physiological data can be used in conjunction with this game. To measure the electrical signal of the heart, we used a wearable ECG device called \textit{Heartypatch} and its recording software\cite{ecg_data_record}. Figure~\ref{Heartbeat_figure} shows a sample ECG recording obtained from the device and its software. 

\begin{figure}[h]
  \centering
  \includegraphics[height=7cm]{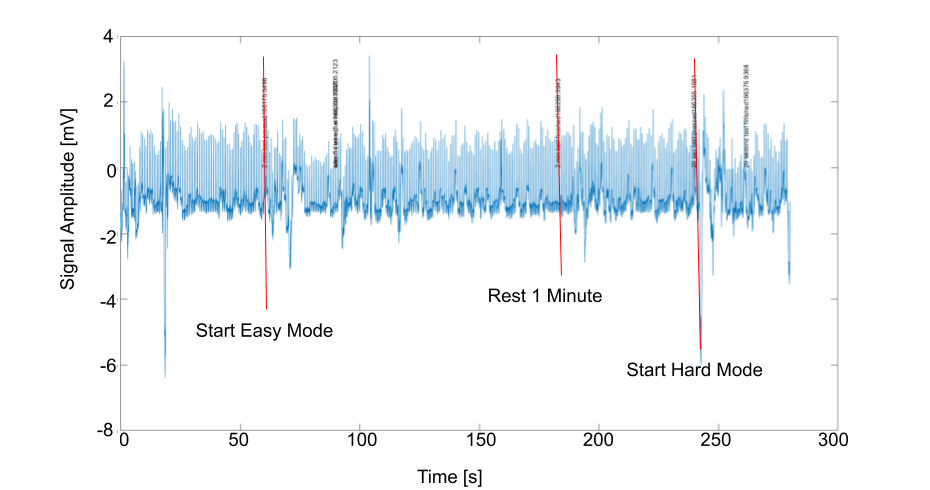}.
  \caption{We recorded the ECG of the player using the HeartyPatch sensor while the player was playing the game, and overlaid it with the event markers. The total recording is four minutes and thirty seconds. For the first minute, the player is resting to normalize their heartbeat. Afterward, for two minutes the player is playing the non-stressful version of the game (easy mode). The player then takes a minute rest and starts the stressful version of the game for twenty seconds (hard mode). 
  \label{ECG_figure}
} 
\end{figure}

The trials we recorded were split into two-minute or twenty-second sessions, with a one-minute break in between sessions. Each session required the player to solve a random puzzle module, from cutting wires to pressing buttons. The two-minute sessions were the easy sessions with little stress, while the twenty-second sessions were meant to be stressful (hard) sessions that tested the player’s ability to solve puzzles under a very short time limit. Figure~\ref{ECG_figure} shows ECG data synchronized with event markers from the game, revealing the various levels of ECG signal depending on how stressful the player was at key points in the game.

\subsection{HRV Measurements}

\begin{figure}[H]
  \centering
  \includegraphics[height=7cm]{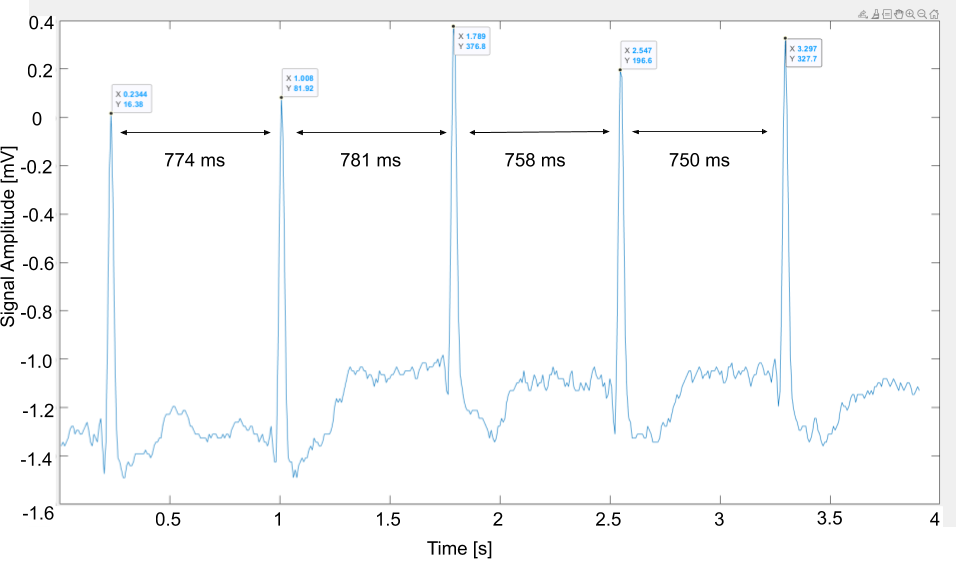}.
  \caption{We recorded the ECG of the player using the wearable HeartyPatch while the player was playing the VR game. The time intervals between consecutive heartbeats are depicted to demonstrate the variability of the time intervals.
} 
  \label{Heartbeat_figure}
\end{figure}

Heart rate is the number of heartbeats per minute. Heart rate variability is the fluctuation in the time intervals between adjacent heartbeats \cite{heartbeat}. To calculate HRV, we first find the time intervals between the consecutive heartbeats using the \texttt{findpeaks} function in MATLAB. Figure~\ref{Heartbeat_figure} shows heartbeats and time intervals between consecutive heartbeats of a sample ECG recording.
We then compute the differences between the adjacent time intervals and then square them.
Finally, we add all those squared differences, divide it by the total number of the squared differences to find the mean, and then take the square root of the mean to find the HRV. 
This process is succinctly shown in the following equations:

\begin{equation}
    y = \sum_{i=1}^{n-1}{(t_i-t_{i+1})^2}
\end{equation}
\begin{equation}
    HRV = \sqrt{y/(n-1)},
\end{equation}
where the $t_i$ is the $i$-th time interval (the difference in time in milliseconds) between two consecutive heartbeats and $n$ is the total number of the time intervals.  

\begin{figure}[H]
  \centering
  \includegraphics[height=6cm]{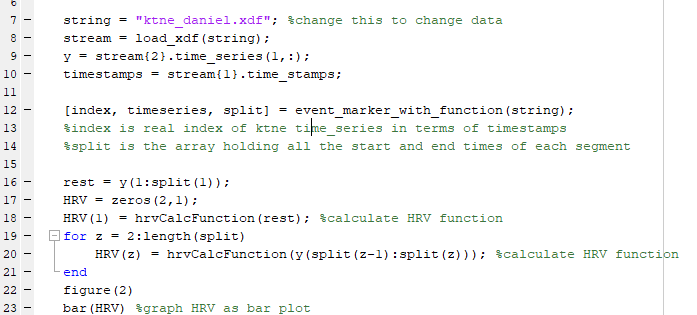}.
  \caption{This MATLAB code takes an xdf data file recorded from LSL and plots the ECG with Event Markers from playing the VR game. It also finds the HRV across multiple sections of the data separated by event markers and plots the HRV values as a bar graph.
} 
  \label{HRV_Code}
\end{figure}

\section{Experimental Results}

This section presents our initial experimental findings that relate the variability in time stress with the physiological responses such as heart rates and HRV. 

\subsection{Experimental Results with HRV Measurements}

The goal of this pilot study was to explore how variability in time stress relates to variability in physiological responses (e.g. heart rate and HRV) of healthy individuals. To this end, we recorded the event markers, the players' ECG, and their behavioral data under three conditions: stressful (hard), non-stressful (easy), and resting conditions. The stressful (hard) condition required the player to solve a puzzle module in twenty seconds, while the non-stressful (easy) condition allowed the player to solve a puzzle module in two minutes and the resting condition allowed the player to take a break for one minute between two sessions. 
In each experiment, the participants played four hard sessions, four easy sessions, and eight resting sessions. The easy and hard sessions were played in random sequence in the experiment. Two healthy individuals participated in this pilot study, resulting in the ECG signal data for a total of eight hard, eight easy, and 16 resting sessions.

After collecting the data through LSL from our experiment, we split the data using the event markers from the game. We identified which sections of the ECG data correspond to easy, hard, or resting sessions. We then analyzed the data from the separated and categorized (easy, hard, or resting) sessions to find the average heart rate and HRV values for the easy, hard, or resting conditions using MATLAB as shown in Figure~\ref{HRV_Code}.

\begin{figure}[H]
    \includegraphics[width=\linewidth]{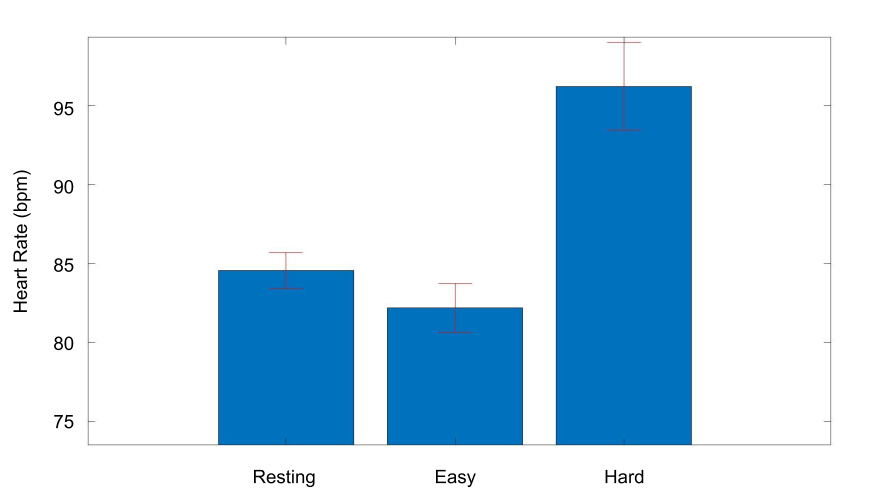}
      \caption{
      This figure shows the average heart rates in beats per minute for two subjects for the 16 resting sessions, eight easy (non-stressful) sessions, and eight hard (stressful) sessions. The error bars denote the variations in the heart rates within the resting, easy and hard sessions.
      }
  \label{heartrate}
\end{figure}

Figure~\ref{heartrate} shows the average heart-rate values under the three (resting, hard, and easy) conditions. The hard (stressful) sessions have the highest average heart rate of 96 bpm. The resting and easy (non-stressful) sessions have average heart rates of 84 and 82 bpm, respectively. The difference between the average heart rates of easy and resting sessions is small, but the error bar for the easy sessions is larger, indicating that the variation of heart rates in the easy sessions is larger. 

\begin{figure}[H]
    \includegraphics[width=\linewidth]{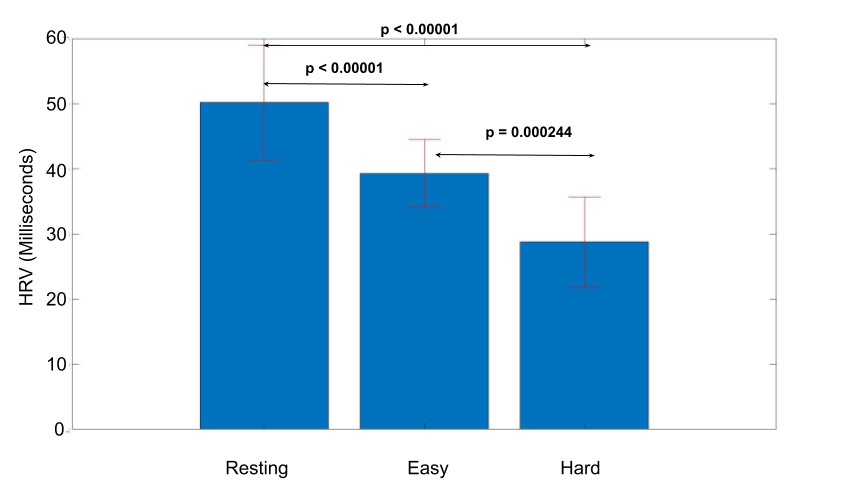}
      \caption{This figure shows the average HRV values under three conditions for two subjects. The statistical testings show that the HRV differences in the resting vs hard, resting vs easy, and easy vs hard are all significant.
      }
  \label{HRV_Values}
\end{figure}

Figure~\ref{HRV_Values} shows the average HRV in milliseconds, along with error bars, under the three (resting, hard, and easy) conditions. As shown in the bar graph, the average HRV value under stressful (hard) condition is around 30 milliseconds while those under the non-stressful (easy) and resting conditions are around 40 milliseconds and 50 milliseconds, respectively. Thus the difference between the average HRV values of the resting and easy conditions is around ten milliseconds, and the difference between those under the easy and hard conditions is also ten milliseconds, which makes the difference between the average HRV values under hard and resting conditions to be about 20 milliseconds. We also note that the average HRV value under stressful (hard) condition is the lowest.

We used statistical tests to examine if the differences in HRV between conditions are statistically significant. To calculate statistical significance, we determined the null hypothesis and an alternate hypothesis. Since the level of statistical significance is often expressed as a $p$-value between zero and one, we calculated $p$-value to show that our results are statistically significant. Typically, a $p$-value less than 0.05 is considered statistically significant. We thus used a one-tailed hypothesis with a significance level of 0.05.

To calculate the $p$-values, we first calculated the standard deviation for both sets of data under the statistical comparison. Then, we calculated the $t$-score and the degrees of freedom using the number of samples in each data set \cite{stat_sig}. Finally, we used a $p$-value calculator to find the $p$-value using the $t$-score and the degrees of freedom \cite{pcalculator}. 

For example, to find the $p$-value between hard (group 1) and easy (group 2) sessions, we used the following steps:  

First, we determined the null hypothesis and the alternative hypothesis as follows:
The null hypothesis: "There is no significant difference in the data sets (HRV values) of group 1 (hard sessions) and group 2 (easy sessions)."
The alternative hypothesis: "There is a significant difference in the data sets (HRV values) of group 1 (hard sessions) and group 2 (easy sessions)."

Second, we calculated the standard deviation $\sigma$ of the HRV values (data sets) for the hard sessions and easy sessions using the following formula: 
\begin{equation}
    \sigma = \sqrt{(\sum_{i=1}^{N}{(x_i - \mu)^2}) / (N-1)},
\end{equation}
where 
$x_i$ is an individual data value in the given group (hard or easy),
$\mu$ is the mean of the data for the given group (hard or easy), and
$N$ is the total sample size of the given group (hard or easy).

Third, we calculated the standard error $s$ between group 1 (hard) and group 2 (easy) sessions using the following formula:
\begin{equation}
    s = \sqrt{(\sigma_1/N_1) + (\sigma_2/N_2)},
\end{equation}
where 
$\sigma_1$ is the standard deviation for the data in group 1 (hard),
$N_1$ is the sample size of group 1 (8 for hard),
$\sigma_2$ is the standard deviation for the data in group 2 (easy),
and $N_2$ is the sample size of group 2 (8 for easy).

Fourth, we calculated the $t$-score using the following formula:
\begin{equation}
    t = |\mu_1-\mu_2|/s,
\end{equation}
where
$\mu_1$ is the mean of the data for group 1 (hard),
$\mu_2$ is the mean of the data for group 2 (easy), and
$s$ is the standard error between group 1 and group 2. 

Next, we calculated the degrees of freedom $d$ using the following formula:
\begin{equation}
    d = N_1 + N_2 - 2,
\end{equation}
where
$N_1$ is the sample size of group 1 (8 for hard),
and $N_2$ is the sample size of group 2 (8 for easy).

Finally, we used the $p$-value calculator \cite{pcalculator} to find the $p$-value based on the $t$-score and the degrees of freedom $d$.

As shown in Figure~\ref{HRV_Values}, the $p$-value for the null hypothesis that there is no significant difference between the average HRV values in hard and easy sessions is $p=0.00244$, which is less than the significance level of 0.05. 

This means that the difference between these values is statistically significant and unlikely to happen by chance. The HRV differences between resting and hard groups and between resting and easy groups are also statistically significant since their $p$-value is much less than 0.05 ($p < 0.00001$).  

Hence, our experimental results showed that the average HRV value is the lowest under the stressful conditions with statistical significance.
 
\subsection{Comparison with Other Results on Heart Rate Variability Under Stressful Conditions}

We compared our findings of the lower HRV values under stress to prior work such as \cite{athleteHR}, which analyzed the influences on HRV Values in athletes, and \cite{Stress_HRV}, which analyzed the relationship between HRV and regional cerebral blood flow. Both studies showed the stress in athletes and higher blood flow was correlated with lower HRV values. 

In the study \cite{hyp}, researchers evaluated the effect of real-life stress on the cardiac response of the subjects. They concluded that stress increases arterial pressure and impairs cardiovascular homeostasis. 

Furthermore, in another study \cite{HRVEffect}, participants performed tasks that either had a physical, mental, or combined load, while their HRV was measured. The study concluded that HRV is affected by changes in physical or mental states, and they were also able to differentiate between resting, physical and mental conditions through the characteristics of HRV.

Researchers surveyed London-based civil servants (aged from 35 to 55) \cite{lowHRV} to evaluate the stress levels due to their work. They reported correlations between high work stress, low physical activity, poor diet, and most importantly lower heart rate variability.

\section{Use of Automatic Speech Recognition as a Tool to Support Communication}

When playing the game, we realized a problem that affected the communication between two players. The player often had to ask the bomb-defusing expert to clarify their instructions multiple times to make sure that they had confidence in what wires needed to be cut or what button sequence needed to be pressed. This process of double-checking resulted in delays and possible confusions while disarming the bomb. This issue becomes more significant for harder modules such as cutting several wires in a sequence where the bomb defuser has to correctly memorize the sequence. 

This section describes how the use of an Artificial Intelligence (AI) assistant in the form of an automatic speech recognition system can help the player to visualize the instructions inside the VR headset. 

\subsection{Use of Speech Recognition During Game Play}

To showcase our concept, we decided to use a readily available speech recognition system in the Unity library called DictionRecognizer. This tool in Unity listens to speech input and returns a string of words it recognizes in the speech input.

\begin{figure}[H]
  \centering
  \includegraphics[height=4.5cm]{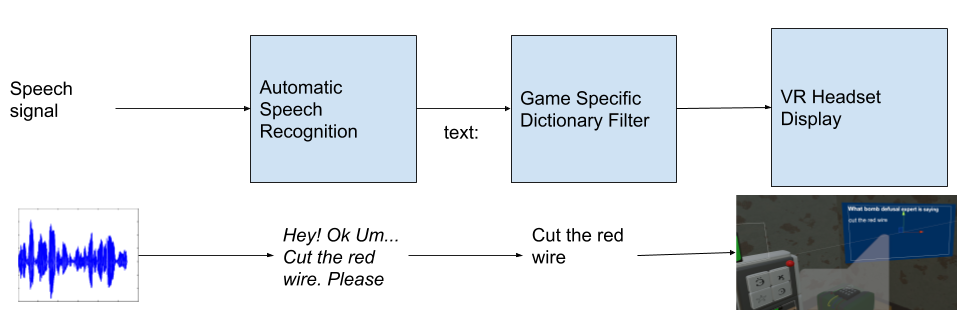}.
  \caption{The game records the speech signal and sends it to the Automatic Speech Recognition program in Unity. The program converts the speech signal to text and sends it through the game-specific dictionary filter, where sentences and phrases that do not contain keywords are cut out. The resulting keywords are displayed in the VR headset.}
  \label{speech_figure}
\end{figure}

To provide just the precise information to the player, we used a dictionary of keywords that are used for solving the puzzle to check if the recognized spoken words are part of the keywords.
This game-specific dictionary filters out unimportant sentences and words to display only keywords such as “wire”, “bomb” or names of colors, so that the player can focus on the important instructions only. We came up with 20 keywords that were important to solve the puzzles and stored them in an arraylist. Each recognized word is then compared against the words that are in the arraylist to determine if they should be shown in the headset display.
Figure~\ref{speech_figure} shows the sequence of this speech-to-text program, the keyword filtering, and how the words appear in the display. 

For time-sensitive tasks such as the tasks in this game, we believe that the use of the speech recognition system with a keyword filter can significantly facilitate the communication between the player who disarms the bomb and the player (the bomb-defusing expert) who gives the instruction.  

\subsection{Recording ECG and Audio While Using the Speech Recognition Tool}

To demonstrate how the speech recognition tool can be used in conjunction with the heart-rate measurements, we developed the above-described speech recognition tool as a feature in the game that can be enabled or disabled while playing the game.

\begin{figure}[H]
  \centering
  \includegraphics[height=7cm]{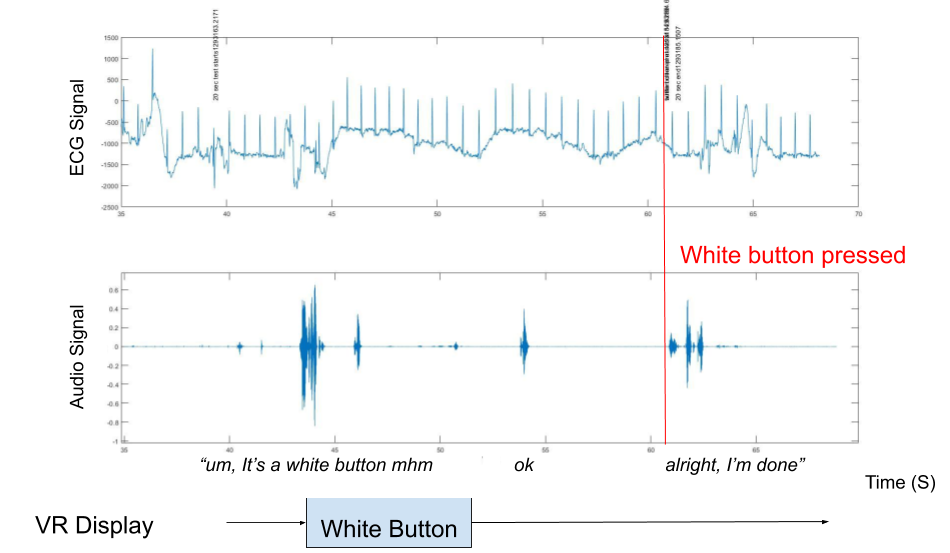}.
  \caption{A sample recording shows the synchronized ECG, audio signals, the output of the AI-based speech-recognition system, and behavioral data during the gameplay. The bomb-defusing expert told the player what to do, "um, it's a white button." The AI-based speech-recognition system detected the keywords, "White button" and displayed it in the VR headset. The player pressed a button at the very end of the allotted time. } 
  \label{audioECG}
\end{figure}  
 
We recorded the player's ECG signals while the player tried to solve the puzzles and disarmed the bomb and at the same time, we also recorded the audio signal of the other player who gives the instructions. Afterward, we overlaid the event markers with both the audio and ECG streams, and synchronized them on the same $x$ axis as shown in Figure~\ref{audioECG}. 

This illustration is intuitive and we expect this tool to improve performance during the gameplay and also believe that it can result in reduced stress which can be measured by the HRV values. Further experiments and studies with multiple subjects are required to quantify our measurements and verify this conjecture.

\section{Conclusions and Next Steps}

In conclusion, we created a VR GWAP framework based on the game "Keep Talking and Nobody Explodes" intending to relate physiological data and stress levels. The framework includes event markers so that the physiological, behavioral, and audio data streams and events in the game are totally synchronized. We used ECG to show how this VR GWAP can be used to assess the physiological correlates of stress. The empirical results were consistent with those reported in prior studies that the HRV values decrease with increasing levels of stress \cite{Stress_HRV, heartbeat, HRVEffect, lowHRV}. 

Our game is meant to be a framework for experimentation so that other researchers can add more tools or sensors. We used ECG to measure heart rate variability as it is known to be correlated with stress. However, this framework can be extended to sensors such as electroencephalogram (EEG) to measure brain activities that are correlated with stress, or eye gaze/pupillometry to find other physiological correlates of stress. This framework also allows a thorough investigation into the speech characteristics of players under stress, for example, to extract features such as pitch or loudness. We believe that this framework can facilitate further research on the connection between physiological signals and stress (and other mental and cognitive states). 

\section*{Acknowledgments}
This work was supported in part by grants from US NSF (CBET-1935860, NCS-1734883, IP-1719130, and SMA-1540943) and US Army Research Lab STRONG Program to TPJ. 
The authors want to thank Robin Xu for helping with the VR game code and Kuanjung Chiang for helping with the Data Analysis in MATLAB.


\bibliography{bibliography}
\bibliographystyle{ieeetr}

\end{document}